\documentclass[aps,prl,twocolumn,groupedaddress,showpacs,amssymb]{revtex4}
\usepackage{graphicx}

\begin{document}

%\preprint{}

\title{$p$-$sd$ shell gap reduction in neutron-rich systems 
and cross-shell excitations in $^{20}$O}

\author{ M. Wiedeking, S.L. Tabor, J. Pavan, A. Volya, 
A.L. Aguilar, I.J. Calderin, D.B. Campbell,  W.T. Cluff, E. Diffenderfer, 
J. Fridmann, C.R. Hoffman, K.W. Kemper, S. Lee, M.A. Riley, B.T. Roeder, 
C. Teal, V. Tripathi, and I. Wiedenh\"over}
\affiliation{Florida State University, Tallahassee, FL 32306, USA} 

\date{\today}

\begin{abstract}

Excited states in $^{20}$O were populated in the reaction 
$^{10}$Be($^{14}$C,$\alpha$) at Florida State University. Charged particles 
were detected with a particle telescope consisting of 4 annularly segmented 
Si surface barrier detectors and $\gamma$ radiation was detected with the  
FSU $\gamma$ detector array.
Five new states were observed below
6 MeV from the $\alpha$-$\gamma$ and $\alpha$-$\gamma$-$\gamma$
coincidence data.  Shell model calculations suggest that most 
of the newly observed 
states are core-excited 1p-1h excitations across the $N = Z = 8$ shell gap.
Comparisons between experimental data and calculations for the neutron-rich 
O and F isotopes imply a steady reduction of the $p$-$sd$ shell gap as 
neutrons are added. 
\end{abstract}


%\keywords{}

\maketitle

The mean field is a central concept in all macroscopic 
many-body phenomena; in nuclei it
leads to shell structure and magic numbers and it provides the 
basis for the most powerful theoretical 
techniques such as the nuclear shell model. In recent
years attention has been focused on light nuclei, which are  
intermediate between macroscopic and 
few-body nucleonic systems, beacuse they can exhibit 
extreme modifications or even complete breakdown of the mean field approach. 
This behavior emphasizes the underlying question of nature of how 
combinations of a few simple building blocks can lead 
to the astonishing diversity and complexity of many-body phenomena.    
Interest in this subject is fueled by experimental advances which allow 
for the exploration of nuclear structure 
at the limit of stability with critical proton to neutron ratios.   

The most significant manifestation of shell breaking is seen in nearly 
semi-magic nuclei, where the mean-field shell structure breaks down as 
a function of the proton to neutron ratio. ``Islands of inversion''
\protect\cite{War90, Cau98} have been
observed around the neutron shell closures at $N=20$ \protect\cite{Pri01} 
and $N=28$ \protect\cite{Cot02}, where 
the mean-field shell structure is weakened due to the 
proton-neutron imbalance  and residual interactions dominate 
\protect\cite{Uts04}. This 
leads to structures where intruder configurations dominate 
low-lying states \protect\cite{Van04}.   

The oxygen isotopes, which lie at the $Z = 8$ shell boundary, provide an 
excellent opportunity to investigate the effects of neutron excess on the 
shell gap and cross shell excitations.   Possible modifications of the
shell structure with neutron excess raise important questions related to
the surprisingly large shift in the neutron drip line between the O and 
F isotopes \protect\cite{Gui89,Sak99}.  Cross shell $p$-$sd$ excitations are 
known to be essential in the structure of nuclei in this region.
Examples include the 110 keV $1^-$ state in $^{19}$F \cite{Ari67} and a 
recent experimental investigation unveiling low-lying intruder 
states in $^{22}$F \protect\cite{Pav04}. 

The availability of suitable reactions has limited the study of the 
neutron-rich O isotopes. Most of the previous work on $^{20}$O
has involved one of the earliest radioactive beams through the $^{18}$O(t,p)
reaction \protect\cite{Pil79,LaF179,You81}.  
Fragmentation reactions were recently used to determine $\gamma$ 
decays from some previously known states \protect\cite{Sta04}. 
Intruder states, which have an odd number of 
nucleons promoted across the $p$-$sd$ shell gap,  
provide the clearest evidence for 
cross shell excitations.  Only 2 intruder states, a $1^-$ at 5.35(10) MeV 
\protect\cite{Try03} and
a tentative (3$^-$) at 5.614(3) MeV \protect\cite{LaF279,Kha00}, had been 
identified in $^{20}$O prior to
the present work.  However, shell model calculations including $p$ shell
excitations predict significantly more states in the region of 4.5 to 6 MeV.
Due to this discrepancy the present experiment was performed to 
study the shell structure of $^{20}$O. 

In this letter, we report on the addition of five new states in 
$^{20}$O below an excitation energy of 6 MeV. With the additional 
experimental information, shell model 
calculations were performed for neutron rich O and F isotopes. These 
provide strong evidence for a reduction of the $p$-$sd$ shell gap 
as neutrons are added.

To populate $^{20}$O, 
a long-lived radioactive beam and target were employed 
through the $^{10}$Be($^{14}$C,$\alpha$) reaction at 21.4 MeV.  The $^{14}$C
beam was provided by the Florida State University (FSU) Superconducting
Accelerator Laboratory.  The 113 $\mu$g/cm$^2$ $^{10}$Be target on a 
1 mg/cm$^2$ thick Pt backing was specially made \protect\cite{Goo74}.
Neutron evaporation dominated the reaction, and $^{20}$O could only be
observed by detecting the $\alpha$ particles.  These were detected in an
E-$\Delta$E Si telescope, consisting of a 150 $\mu$m $\Delta$E detector
and a stack of three 1500 $\mu$m E detectors.  All four Si wafers were divided
into 4 annular segments and placed at $0^\circ$ relative to the beam.
A 32 mg/cm$^2$ thick gold foil was placed between the target and detector
telescope to stop the beam. $\gamma$ radiation was detected in coincidence
with the $\alpha$ particles using the FSU Compton suppressed 
$\gamma$-detector array. During the experiment $4.3 \times 10^6$ 
$\alpha$-$\gamma$ events were recorded. 

The data were analyzed using GNUSCOPE \protect \cite{Pavphd}.  The $\alpha$
spectra from the 5 annular segments were kinematically corrected, and
a Doppler correction was applied to the $\gamma$ spectra.  Although the
target contained significant amounts of $^{16}$O and $^{12}$C, the
corresponding reaction products, $^{26}$Mg and $^{22}$Ne, are well known
and their lines were removed from consideration.

\begin{figure}
\includegraphics[scale=0.35]{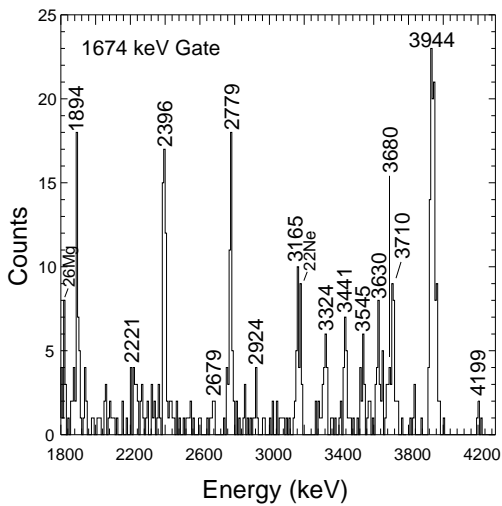}
\caption{\label{fig:1674gate} Transitions in coincidence with $\alpha$ 
particles and the
1674 keV $\gamma$ ray ($2^+_1 \rightarrow 0^+_1$). 
The lines in this spectrum marked by their energies correspond to decays 
in  $^{20}$O.}
\end{figure}

\begin{figure}
\includegraphics[scale=0.4]{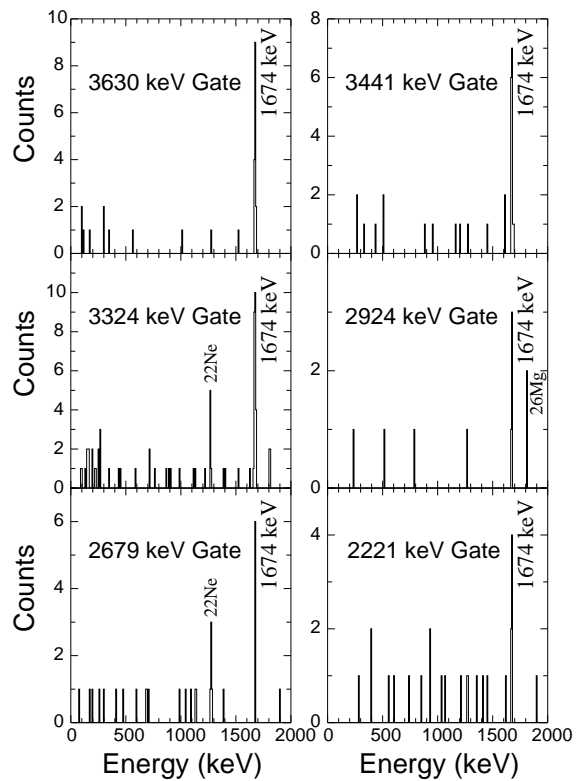}
\caption{\label{fig:NewGates} Spectra in coincidence with $\alpha$ particles 
and newly observed transitions in $^{20}$O.}
\end{figure}

\begin{figure}
\includegraphics[scale=0.4]{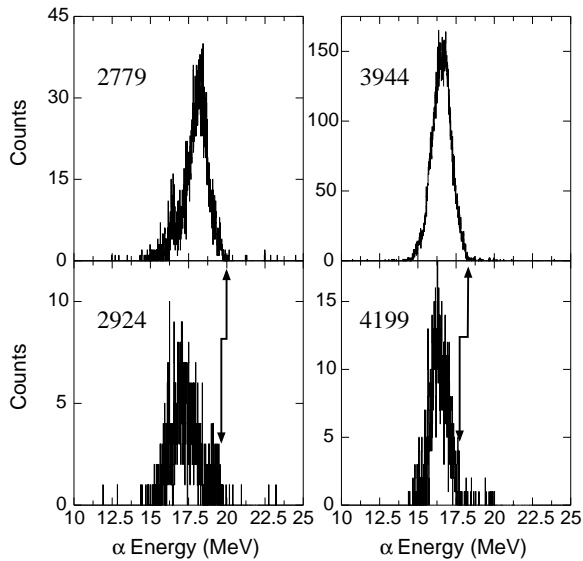}
\caption{\label{fig:Cutoff2} $\alpha$ spectra   
for two $\gamma$-ray pairs originating from states with similar 
excitation energies. 
The arrows indicate the cutoff location.}
\end{figure}

\begin{figure}
\includegraphics[scale=0.3]{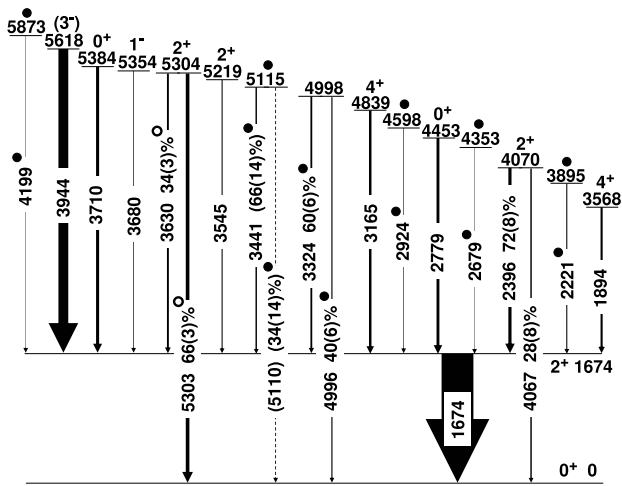}
\caption{\label{fig:20Ols} The level scheme as deduced from current and previous work. 
The widths of the lines indicate their intensities normalized to the 
1674 keV transition. Branching ratios are given where applicable. 
States and transitions marked with solid circles are new additions. 
$\gamma$ transitions with open circles were tentatively reported in 
Ref. \protect\cite{You81} and are confirmed in the present work.}
\end{figure}

A portion of the $\alpha$-$\gamma$-$\gamma$ spectrum in coincidence with the
lowest $2^+_1 \rightarrow 0^+_1$ transition at 1674 keV is shown in 
Fig.~\ref{fig:1674gate}. 
The $\alpha$-$\gamma$-$\gamma$ coincidences with most of the newly assigned 
lines
are shown in Fig.~\ref{fig:NewGates}.  Clear coincidences with the 1674 keV 
transition 
($2^+_1 \rightarrow 0^+_1$) can be seen in all these spectra.  Placement of 
the 
$\gamma$ transitions was verified by the coincident $\alpha$ energies, as 
described 
in Refs. \protect\cite{Coo02, Hof03}. Some $\alpha$ spectra obtained 
by gating on $\gamma$ lines are shown in Fig.~\ref{fig:Cutoff2}.   
An extensive search for possible contaminants 
was performed for all transitions before a firm assignment to $^{20}$O 
was made. 
The level scheme from this and previous work is shown in 
Fig.~\ref{fig:20Ols}. The uncertainties in $\gamma$-ray energies range 
from 1 keV up 
to 6 keV for the weaker new transitions.

Although the 5 new states at 3895, 4353, 4598, 5115, and 5873 keV have not
been previously reported, a careful examination of previous studies of $^{20}$O
does reveal hints of their existence.  Very small peaks are visible in
the $^{18}$O(t,p) spectrum shown in Fig. 1 of Ref. \protect\cite{LaF279} 
at 3.9, 4.6, 
5.1, and possibly 5.9 MeV.  While examination of only one spectrum at one 
angle cannot
rule out contaminants, the weakness of these peaks is consistent with their
weak population in the present experiment.  Interestingly, a tentative state
was reported at 5.83(20) MeV in an early paper \protect\cite{Hin62} and 
included
in an earlier compilation \protect\cite{Ajz78}.  A state was reported at 6.02
MeV in another measurement \protect\cite{Pil79}.
The 5.83 MeV state was dropped from the subsequent compilation 
\protect\cite{Ajz83},
possibly because it was not confirmed in Ref. \protect\cite{LaF279}.  The
state observed here at 5873 keV may correspond to one or both of these 
previously reported states.

While shell model calculations restricted to the $p$ shell or to the $sd$
shell have been very successful in describing the structure of nuclei in these
regions, the description of excitations between the $p$ and $sd$ shells
has been much more limited.  The lack of experimental data and 
increased dimensionality of such calculations together with the center-of-mass 
issue \protect\cite{Glo74} resulted in the lack of well tested interactions. 
To gain an understanding of the current experimental results, $p$-$sd$ shell 
model calculations were performed for neutron rich O and F isotopes  
using and slightly modifying (as discussed below) cross-shell interactions 
from Refs. \protect\cite{McG73,War90}. 
Calculations were also conducted taking excitations into the $fp$ 
shell into account. However, at excitation energies of interest 
$fp$ configurations proved to be of no importance. 

\begin{figure}
\includegraphics[scale=0.5]{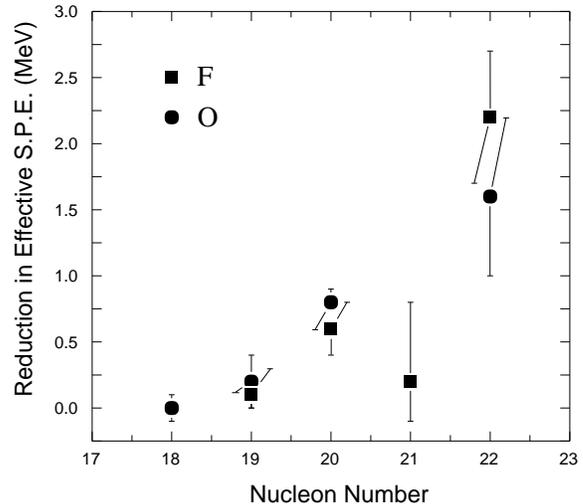}
\caption{\label{fig:psdgapmod} Systematic reduction of the $p$-$sd$ shell 
gap as neutrons are added to O and F nuclei. Shell model caculations 
were performed and the shell gap adjusted based on the comparison 
to available experimental data from the current work and Refs. \cite{Pav04,
Cor04, endsf}. The error bars reflect  the current experimental 
knowledge available for the respective nuclei.}
\end{figure}

For the calculations, an m-scheme shell model code developed 
at FSU \protect\cite{Vol04} was used  
which allows for a full diagonalization of the $p$-$sd$ shell simultaneously. 
However, the interactions based on a perturbative
merging of the $sd$ and $p$ shells imply a particle-hole hierarchy.     
Establishing a trend in reproducing the   
experimental data is a key step in validating the model calculations.
The monopole corrected single-particle energies in the $sd$ shell 
were adjusted to reproduce single-nucleon excitations in $^{17}$O. Thus, the 
model is in good agreement with USD shell model results. 
The cross shell excitations in $^{16}$O are also well reproduced. The 
perturbative combined $p$-$sd$ shell model essentially has one parameter, 
the shell gap.  The role of this gap becomes more essential in
establishing the relative positions of cross shell excitations as the 
neutron number $N$ increases above the shell closure at $N=8$. 
Considering negative-parity states in $^{18,19,20,22}$O and using the original 
interaction, a significant trend of deviation is found. For example  
the 1p-1h states in $^{22}$O are predicted to be more than 1 MeV above the 
first tentatively observed negative-parity state at 5.8 MeV 
\protect\cite{Cor04}. The agreement with experimentally observed states 
can be improved significantly by reducing the $p$-$sd$ shell gap with 
increasing in neutron number.
The necessity to reduce the shell gap to obtain the best agreement between 
experimental and theoretical states is also observed in the neutron-rich 
$^{19,20,22}$F isotopes. This  suggests a general $p$-$sd$ 
shell gap decrease as 
more neutrons are added, in agreement with theoretical predictions 
\cite{Bro}. The shell gap dependence on neutron number 
for O and F isotopes is shown in  
Fig.~\ref{fig:psdgapmod}. The diminishing of the shell gap emphasizes 
the approach to a region of intruder dominated structures. 

\begin{figure}
\includegraphics[scale=0.43]{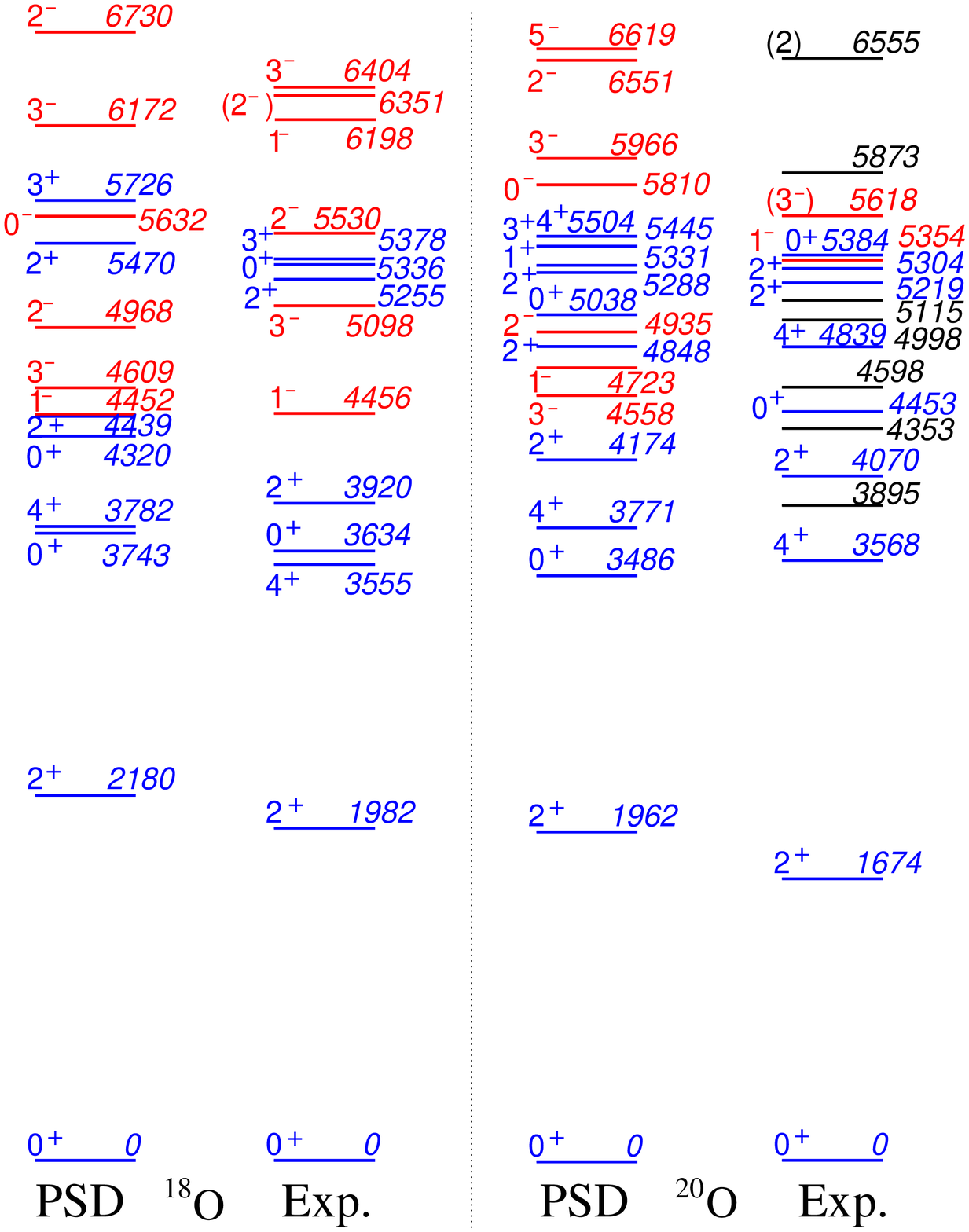}
\caption{\label{fig:psdexpaper} (Color online) Comparison of calculated and 
experimentally observed states in $^{18}$O and $^{20}$O. States in blue 
indicate positive-parity states (0p-0h and 2p-2h), states in red are 
negative-parity states (1p-1h). The levels in black are new additions, 
except the 6555 keV state in $^{20}$O where no parity assignment is 
available. For $^{18}$O the standard $^{16}$O closed shell single-particle 
energies are used, whereas for $^{20}$O the $p$-$sd$ 
shell gap was reduced by 850 keV.}
\end{figure}

The agreement with experimentally observed states in $^{20}$O can be improved 
significantly by reducing the $p$-$sd$ shell gap by 850 keV.
Fig.~\ref{fig:psdexpaper} compares the calculated and experimentally 
observed states 
in $^{18}$O and $^{20}$O, where the  closed shell $^{16}$O
single-particle energies are used for $^{18}$O and a reduced $p$-$sd$ 
shell gap 
of 850 keV is used for $^{20}$O.
The theoretical states shown in blue are mainly the pure 
$sd$ ones involving only 
valence neutrons outside the closed $^{16}$O core using the USD interaction. 
Two states at 3743 and 5470 keV in $^{18}$O and at 3486 and 4848 keV 
in $^{20}$O are due to 2p-2h excitations.
Previously known experimental  positive-parity states are also shown in blue.  
There is good agreement for the positive-parity states in $^{18}$O and 
$^{20}$O.  
The tendency to overpredict the
energy of the first $2^+$ state is somewhat greater in $^{20}$O. 
Negative-parity 1p-1h states (red) are predicted to start at about 4.5 MeV in
both nuclei.  There are good candidates in $^{18}$O within a few hundred keV
for all but one of the predicted negative-parity states.  In $^{20}$O
the lowest previously established negative-parity state 
lies about 800 keV above 
the first predicted one.  However, some of the newly observed levels lie
reasonably close to a number of the predicted negative-parity states.
Unfortunately, the newly observed
states are populated too weakly to determine their spins or parities, but 
there is a good qualitative agreement between experiment and
theory in $^{20}$O with the addition of the new states.  

Further experiments probing whether
inversion can be reached before the drip-line are of great interest. 
An analogous situation appears one shell below, in the $Z=2$ He 
isotopes, where the intrusion of $sd$ excitations in $p$-shell configurations 
becomes important in heavy helium isotopes. 
Although the $^8$He dripline is reached
before inversion, the resonant states exhibit significant fragmentation 
of the single particle decay strength due to $sd$-shell intruder configurations 
\cite{Rog03}. It is still controversial if the lowest energy resonance 
corresponding to unbound $^9$He is an inverted structure with
an odd-particle in the $sd$ shell and thus of positive parity. 
The intruder ground state in $^{11}$Be \protect\cite{Ajz83} is another 
example of a weakened shell gap. 
     
To conclude, in this work we report five new states that 
were observed in semi-magic $^{20}$O below 6 MeV through
the measurement of $\alpha$-$\gamma$ and $\alpha$-$\gamma$-$\gamma$
coincidences following population in the $^{10}$Be($^{14}$C,$\alpha$)
reaction using a long lived radioactive $^{14}$C beam and a specially
prepared radioactive $^{10}$Be target.  Many of the newly observed levels
appear to be the predicted core-excited 1p-1h negative-parity 
states. Comparison between shell model calculations and experimental 
observation of neutron rich O and F isotopes implies a systematic reduction 
in the effective $p$-$sd$ shell gap, indicating a weakening of the 
shell structure as neutrons are added. 

\begin{acknowledgments}

This work was supported in part by the U.S. National Science Foundation under 
grant No. 01-39950 and the state of Florida. The authors thank 
D. Robson for helpful discussions.

\end{acknowledgments}

\end{document}